# Coupling of spontaneous emission from GaN/AlN quantum dots into silver surface plasmons

**Arup Neogi**

*Department of Physics, University of North Texas, Denton, TX 76203*

**Hadis Morkoç**

*Department of Electrical Engineering, Virginia Commonwealth University, Richmond, VA 23284, USA*

**Takamasa Kuroda and Atsushi Tackeuchi**

*Department of Applied Physics, Waseda University, Shinjuku, Tokyo, Japan*

We have demonstrated surface-plasmon induced change in spontaneous emission rate in the ultraviolet regime at ~ 375-380 nm, using AlN/GaN quantum dots (QD). Using time-resolved and continuous-wave photoluminescence measurements, the recombination rate in AlN/GaN QD is shown to be enhanced when spontaneous emission is resonantly coupled to a metal-surface plasmon mode. The exciton recombination process via Ag-surface plasmon modes is observed to be as much as 3-7 times faster than in normal QD spontaneous emission and depends strongly on the emission wavelength and silver thickness.

*OCIS codes: Surface plasmons, nitride semiconductors, quantum dots, spontaneous emission, and photoluminescence.*

Surface plasmon polariton (SPP) optics is a potentially attractive approach to photonic integration [1,2]. The electromagnetic properties related to the electron plasma effects are significantly different from the properties of ordinary dielectric materials in the frequency range below the plasma frequency where the real part of a dielectric constant is negative. In this frequency range the wave-vector of light in the medium is imaginary, and therefore there is no propagating electromagnetic modes in such a medium. Different surface effects and non-local (spatial dispersion) effects in noble metals can contribute to corrections to the surface plasmon (SP) frequency [3, 4]. Thin metallic films containing nanoscale surface features results in giant enhancement of linear and nonlinear optical responses. These enhancements are associated with SP or SPP modes whose characteristics are strongly dependent on the geometric structure of the metallic component of the medium and can be further enhanced via the nonlinearities in the material system.

SPP optics is recently being extensively applied to alter the intrinsic properties of localized light emitters and sensors [5-7]. The concept of modifying the spontaneous emission (SE) rate of radiating dipoles within a cavity was first predicted by Purcell [8] in atomic physics and is being rapidly extended to semiconductor microcavities to develop a new class of optoelectronic devices [1-5]. An ability to enhance the SE rate of a solid state emitter or excitonic recombination in the weak coupling regime would in particular allow the fabrication of ultrafast detectors and/or high efficiency light-emitting diodes [2]. The SE can be modified by altering the photonic density of states by resonantly coupling light to a structure such as a metallic film, which exhibits surface plasmon resonance. Such a system can be realized by a single quantum well or a quantum dot located near a thin metal film [6, 7, 10,11].

A single quantum well (QW) or quantum dot (QD) can experience strong quantum electrodynamical coupling to a SP mode if placed within the SP fringing field penetration depth. In this regime, an electron-hole pair in the confined nanostructure recombines and emits a photon into a SP mode instead into free space. The degree of SE rate modification for a given wavelength depends on the SP density of state (SP-DOS) at that wavelength. The strongest enhancement occurs near the asymptotic limit of the SP dispersion branch, the SP "resonance" energy $E_{sp}$, where the SP DOS is very high. Non-resonant, SP-mediated SE enhancements as large as 6 have been observed from GaAs QWs near thin Au films. [12]. Even higher enhancements are possible for wide bandgap semiconductors whose emission wavelength is coincident with $E_{sp}$ and also due to the higher photonic-DOS. SE into surface plasmons was observed to be 55 times faster than normal SE from InGaN quantum wells [11] within the visible wavelength range. The SE into the SPs estimated from the time-resolved photoluminescence measurement was ~ 100 times faster than the normal SE from the InGaN QW [6]. However, modifying the spontaneous emission rate via SP interaction in the ultraviolet range is challenging as the emission energy is above the asymptotic SP energy branch, resulting in a reduction in the coupling between the plasmon modes and the higher energy localized emitters.

In this report, time-resolved PL (TRPL) measurements of a partially silver-coated AlN/GaN QD directly demonstrate the SP-mediated exciton recombination from QDs in the ultraviolet regime. The effect of resonant and off-resonant optical excitation on surface-plasmon mediated electron-hole recombination process in the QDs has also been investigated using time-resolved PL spectroscopy.

The SP energy of Ag (with bulk plasmon energy ~ 3.76 eV) is modified (at the interface ~ 2.92 eV) by the GaN dielectric constant as $\varepsilon'_{Ag}(\omega) + \varepsilon'_{GaN}(\omega) = 0$, where $\varepsilon'_{Ag}(\omega)$ and $\varepsilon'_{GaN}(\omega)$ are the real parts of dielectric constant. The proximity of the SP energy of Ag and bandgap of GaN (bulk bandedge ~ 3.45 eV) is rather fortuitous for enabling resonant energy transfer due to coupling of free-electron or exciton in semiconductors to SP modes at the metal-semiconductor interface. Ag-GaN based semiconductor nanostructures are ideal for developing plasmonic nanostructures for optoelectronic devices in the UV-visible wavelength.

GaN dots on AlN spacer layers were grown on sapphire substrate by molecular beam epitaxy (Fig.1.). Twenty period QD were grown on a strain engineered AlN/GaN buffer layer by conversion of sprayed Ga to GaN by nitridation in an atmosphere of ammonia. QD size varies significantly depending whether the QDs are allowed to evolve under vacuum (before covering with AlN), or not, as a result of a ripening mechanism. The underlying bulk GaN buffer layer is more than 50 nm from the surface and is not influenced by the SP modes at the metal-semiconductor interface. The dot size of GaN/AlN nanostructures can be tailored such that the photoluminescence emission from the QDs is resonant to the Ag-SP energy. In this article, two different types of QD samples a relatively larger QD sample is investigated where exposure of the QDs to vacuum for about 1 minute increased their average height to 7-9 nm (35 nm diameter) as a result of ripening mechanism [13].

In Figure 2, we compare the room temperature emission and absorption characteristics of the strained GaN QDs respectively from photoluminescence (PL) and photoluminescence excitation (PLE) spectra. Continuous-wave PL measurements were performed using He-Cd laser excitation at 3.81 eV. GaN QD exhibits strong PL emission with a peak 3.3 eV, i.e., 140 meV below the underlying bulk GaN energy gap at 3.43 eV. The red-shift in the PL emission due to the QD size is due to the presence of a giant piezoelectric field in the QDs along the c-axis exceeding 2 MV/cm [13-15]. The PLE spectra measured using a 150 W Xe-lamp shows a broad absorption peak from QD clusters at 3.62 eV. This PLE peak, which is blue, shifted by 322 meV compared to the PL emission in the QDs reveals that the absorption occurs in relatively smaller sized QDs. The carriers relax to the relatively larger sized dots, which emits at wavelength below the bulk GaN bandgap energy. The resulting built-in strain also contributes to the large Stokes shift in the emission from the QDs. The variation in dot size distribution was uniform throughout these QD samples.

To modify the exciton recombination rate in QD via SP coupling, the QDs were partly coated with a layer of Ag film (8 nm thick) using thermal vapor deposition. The SP fringing field penetration depth into GaN semiconductor is given by [11] $Z = \lambda/2\pi[\varepsilon'_{Ag} + \varepsilon'_{GaN}/\varepsilon'_{GaN}{}^2]^{1/2} = 40$ nm for Ag on GaN at 2.92 eV, where $\varepsilon'_{GaN} = -\varepsilon'_{Ag} \sim 6$. The GaN buffer layer is beyond z (> 40 nm from the surface) and is not coupled to the Ag-SP modes. The emission from QD is normalized with respect to the emission from the PL from the underlying bulk GaN layers to estimate for the reflection losses due to the silver metal layer.

Figure 2 also shows the PL spectrum from silver-coated GaN/AlN QDs. The PL intensity from the silvered side reduces by 3-5 times at the peak QD PL spectrum compared to the unsilvered side, which indicates a corresponding enhancement in the SE rate in the presence of the Ag-film. The steady-state PL spectrum can be explained with the simple empirical formulation of the Purcell enhancement factor by considering the non-radiative processes to be insignificant compared to the enhanced SE into the plasmon modes in the Ag- coated part.
A more direct demonstration of enhanced exciton recombination, free from the assumptions about silver reflectivity and absorption, involves comparative measurement of the luminescence decay rate from the photoexcited SQW on the silvered and unsilvered sides. The ratio of the radiative PL decay rate from the uncoated and Ag-coated parts of the sample is a direct measurement of the recombination rate enhancement. This ratio, which is an analog of the Purcell [8] factor $F_p$, can be expressed as simply as room temperature TRPL measurements were performed using frequency tripled optical pulses generated from a 100 MHz Kerr-lens mode-locked Ti: Sapphire laser with average incident pump power ($I_p$) of 10 mW (~ 1.8 $\mu J/cm^2$). The pump excitation energy was 267 nm (4.64 eV), The luminescence signal was dispersed in a 1200 ln/mm grating spectrometer and measured simultaneously using a Hamamatsu streak camera with a resolution of 20 ps.

Figure 3a shows the PL decay measured at the bulk GaN bandedge energy (3.43 eV). As the GaN buffer layer in beyond the SP field fringing depth, the PL emission energy from the 3D GaN buffer layer is above the Ag-plasmon energy (3.76 eV), there is no observed change in the PL intensity decay time in the absence of coupling of the SP modes to the emission arising from electron-hole recombination within the bulk GaN layer.

The effect of surface-plasmon coupling to the excitonic emission from GaN quantum dots was studied by the comparing the carrier recombination rate at the peak PL emission energy from QDs at 3.3 eV (Fig.3b). It is observed that due to the nonradiative recombination there is a biexponential decay from both the silvered as well as the nonsilvered side. However, the fast decay component from the silvered side is twice as fast compared to nonsilved side due to presence of the stimulated emission of polaritons via the plasmon modes. The silver nanolayer on the rough surface of the QD layers presumably act as a radiating dipole antenna and the emission from the QD are resonantly coupled into the plasmon modes satisfying the momentum conservation in a particular direction. The slower decay component from the silvered side is about 3-5 times faster compared to the normal QD emission at 3.3 eV and to the off-resonant GaN buffer layer emission at 3.43 eV. The reduction is the carrier recombination time corresponds to the decrease in PL intensity from the silvered side. The difference in the luminescence between the QD and bulk GaN layer increases with the thickness or diameter of the silver nanoshell structures. In Al(Ga)N/GaN QWs, no change in the PL intensity is observed in the presence of silver nanostructures due to lack of directional emission in a 2-D planer QW nanocavity, unlike in QDs where the dipole emitters in one or more QDs are likely to be oriented to the P-DOS. The enhancement in QDs depends on the thickness of silver nanolayer at the surface and its distance from the QD layer.

Figure 4 plots the Purcell factor ($F_{sp} = 1 + t_{QD}/t_{Ag-QD}$

## Conclusions

We have demonstrated the ability to enhance the SE rate in Ag-coated GaN/AlN QDs due to resonant SP interaction in the ultraviolet wavelength range at 3.3 eV. The enhancement of the SE rate is proportional to the quenching of PL intensity in the QD structures.

## *References*

[1] M. Ohtsu, Proc. SPIE–Int. Soc. Opt. Eng., **4416**, . 1, (2001).

[2] M. Borditsky, R. Vrijen, T.F. Krauss, R. Coccioli, R. Bhat and E. Yablanovich, J. Lightwave Tech. **17**, 2096 (1999).

[3] T.W. Ebbesen, H.J Lezec,, H.F. Ghaemi, T. Thio and P.A. Wolff Nature **391**, 667 (1998).

[4] H.J.Lezec, A.Degiron, E.Devaux, R.A.Linke, L.Martin-Moreno, F.J.Garcia-Vidal, T.W. Ebbesen Science **297**, 820(2002).

[5] T. Kawazoe, M. Ohtsu. Appl. Phys. Lett. **82**, 2957 (2003)

[6] A. Neogi, C.W. Lee, H. Everitt, E. Yablanovich, Phys. Rev. B **66**, 153305 (2002)

[7] J.Gerard, B. Sermage, B. Gayral, B. Legrand, E. Costard E and V. Tierry-Mieg V. Phys. Rev. Lett. **81**, 1110 (1998).

[8] E.M. Purcell, Phys. Rev., **69,** 681 (1946).

[9] K.H. Drexhage , (1974). in: Progress in Optics XII (Wolf, E., Ed.), North Holland, Amsterdam.

[10] M. Xiao, Jia-Yu Zhang, and Xiao-Yong Wang, Proc. Int. Quantum Electronics Conference, **IFE 7**, San Francisco (2004)

[11] I. Gontijo, M. Boroditsky, E. Yablonovitch, S. Keller, U. K. Mishra, and S. P. DenBaars, Phys. Rev. B., **60**, 11564 (1999).

[12] H.E. Hecker, R. A. Hopfel, N. Sawaki, T. Maier and G. Strasser, Appl. Phys. Lett.**75**, 1577, (1999).

[13] H. Morkoc, A. Neogi, and M. Kuball, Mat. Res. Soc. Proc. Vol. **749**, T6.5.1/N8.5.1/Z6.5.1, (2004).

[14] A. Neogi, H.O. Everitt, H. Morkoc , IEEE Trans. in Nanotechnolgy, **3**, (2004) {In Press}.

[15] F.Widmann, B. Gayarel, Phys. Rev. B **58**, R15989 (1998).

Figure Caption

Fig.1. Structure of silver-GaN Quantum dot material system for the enhancement of spontaneous emission rate

Fig.2. Normalized photoluminescence spectra from GaN QDs and silver covered GaN QDs.. Quenching of PL intensity in Ag-film covered GaN QDs due to resonant surface plasmon coupling This figure also shows the PLE spectrum of the GaN QDs.

Fig.3. Comparison of PL decay characteristics from GaN QDs and Ag-coated GaN QDs
  (a) at bulk GaN emission energy (~ 3.43 eV)
  (b) Comparison of PL decay characteristics at bulk peak GaN QD emission energy (~ 3.3 eV)

Fig.4. Surface-plasmon-induced spectral luminescence dip, plotted as a Purcell enhancement factor $F_p(\omega)$ for radiation into surface-plasmon modes rather than into external electromagnetic waves.

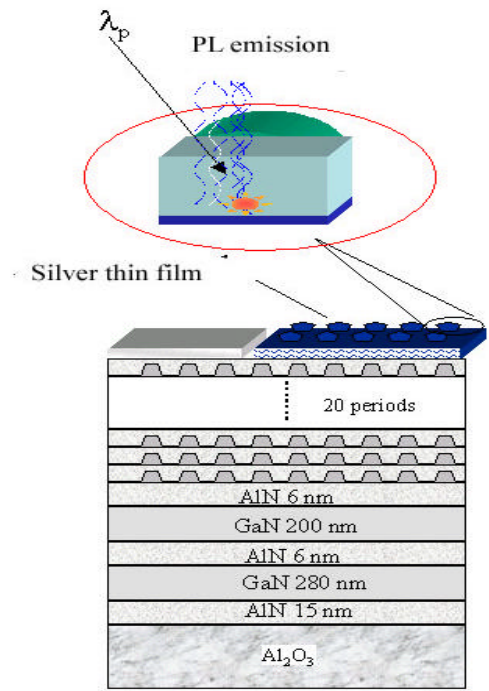

Fig.1.

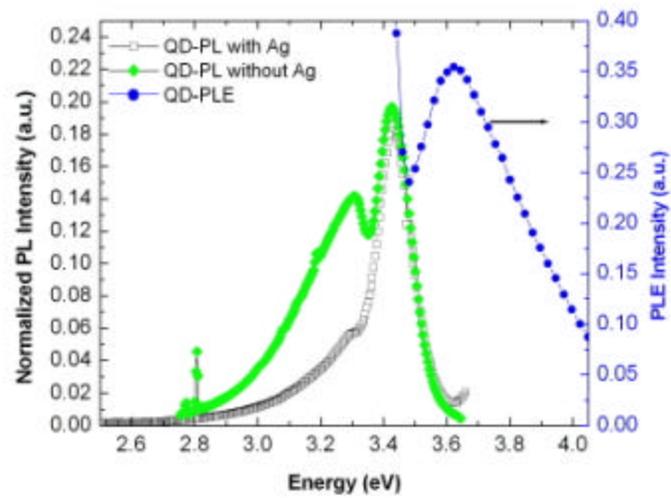

Fig.2.

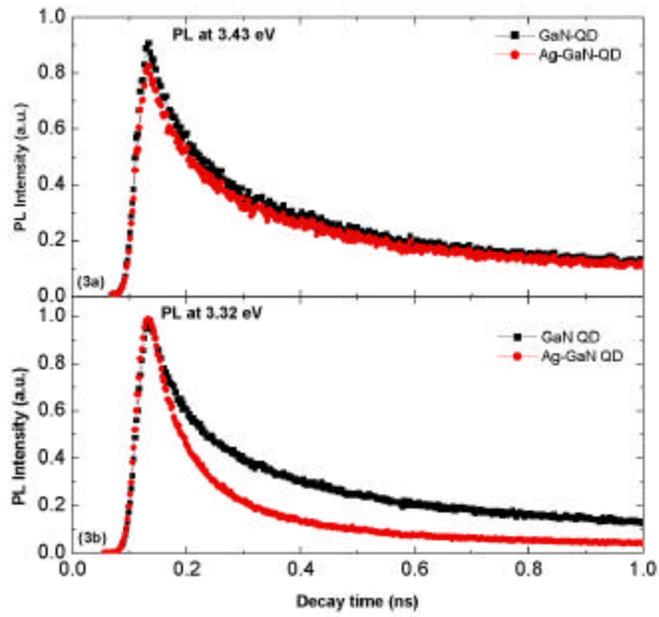

Fig.3.

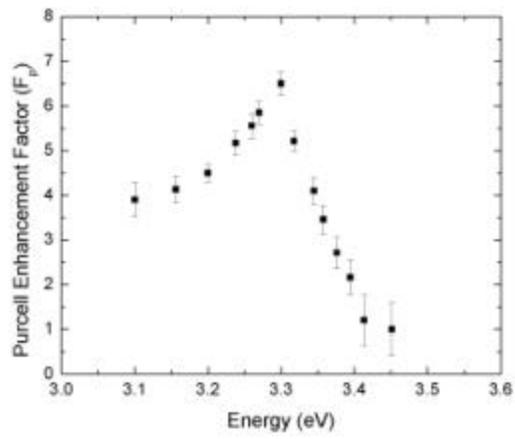

Fig.4.